\documentclass[twocolumn,showpacs,preprintnumbers,amsmath,amssymb]{revtex4}

\usepackage{graphicx}
\usepackage{dcolumn}
\usepackage{bm}

\def\TL{\hfil$\displaystyle{##}$}
\def\TR{$\displaystyle{{}##}$\hfil}

\def\fixit#1{}

\def\overleftrightarrow#1{\vbox{\ialign{##\crcr
     $\leftrightarrow$\crcr\noalign{\kern-0pt\nointerlineskip}
     $\hfil\displaystyle{#1}\hfil$\crcr}}}

\def\lsim{\mathrel{\mathstrut\smash{\ooalign{\raise2.5pt\hbox{$<$}\cr\lower2.5pt\hbox{$\sim$}}}}}
\def\gsim{\mathrel{\mathstrut\smash{\ooalign{\raise2.5pt\hbox{$>$}\cr\lower2.5pt\hbox{$\sim$}}}}}

\def\sqr#1#2{{\vcenter{\vbox{\hrule height.#2pt
         \hbox{\vrule width.#2pt height#1pt \kern#1pt
            \vrule width.#2pt}
         \hrule height.#2pt}}}}

\def\href#1#2{#2}

\def\lbldef#1#2{\expandafter\gdef\csname #1\endcsname {#2}}
\def\eqn#1#2{\lbldef{#1}{(\ref{#1})}%
\begin{equation} \eqalign{#2} \label{#1} \end{equation}}
\def\eqalign#1{\vcenter{\openup1\jot
    \halign{\strut\span\TL & \span\TR\cr #1 \cr
   }}}

\begin{document}

\preprint{PUPT-2098}

\title{Degenerate eigenvalues for Hamiltonians\\ with no obvious symmetries}

\author{Steven S. Gubser}
\author{Robert K. Bradley}
\affiliation{Joseph Henry Laboratories, Princeton University, Princeton, NJ 08544}

\date{\today}

\begin{abstract}
 Certain Hamiltonians based on two coupled quantum mechanical spins exhibit degenerate eigenvalues despite having no obvious non-abelian symmetries.  Operators acting to permute the degenerate states do not have a simple form when expressed as polynomials of the generators of rotations for the respective spins.  As observed in \cite{HapperEtAl}, one such Hamiltonian helps explain resonances in the spin relaxation rate of optically pumped ${\rm Rb}_2$, as a function of applied magnetic field.  We give an explanation of why the degeneracies exist, based on properties of the commutator and anti-commutator of the Hamiltonian and its image under magnetic field reversal.
\end{abstract}

\pacs{03.65.Fd, 31.15.-p, 11.30.Pb, 02.10.Ud}

\maketitle

\section{Introduction}
\label{Introduction}

\begin{figure}[t]
  \vskip20pt
  \centerline{\scalebox{0.75}{\includegraphics{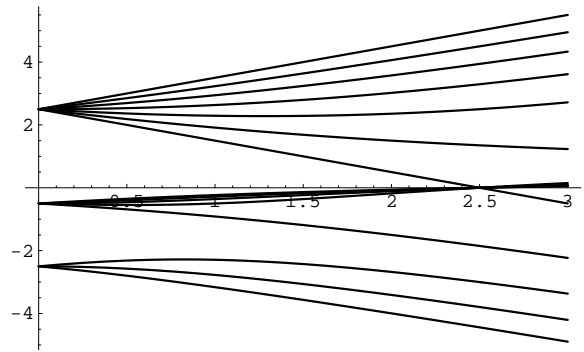}}}
  \vskip0pt
  \caption{Eigenvalues of $H(b)$ plotted against $b$, for $k=2$ and $s=1$.  The degeneracy of interest is at $b=2.5$.}\label{figA}
 \end{figure}

Consider a system composed of two quantum mechanical spins, one of them spin $k$ and the other spin $s$, where we assume $s \leq k$.  The angular momenta for the spins are $\vec{K} = (K_x,K_y,K_z)$ and $\vec{S} = (S_x,S_y,S_z)$, where the components are Hermitian operators acting, respectively, on the $(2k+1)$-dimensional or $(2s+1)$-dimensional Hilbert spaces for spin $k$ or spin $s$.  The Hamiltonian
 \eqn{HapperH}{
  H(b) = \vec{K} \cdot \vec{S} + 1/2 + b S_z
 }
has, for $b \neq 0$, no obvious symmetries other than rotational symmetry around the $z$-axis.  More precisely, $H(b)$ commutes with $J_z$, where $\vec{J} = \vec{K} + \vec{S}$ is the total angular momentum, but there is no other obvious combination of the components of $\vec{K}$ and $\vec{S}$ with which $H(b)$ commutes.  Yet, for $s=1$ and any $k$, a glance at the spectrum of $H(b)$ as a function of $q$ reveals a $(2k+1)$-fold degeneracy at zero energy when $q=k+1/2$: see figure~\ref{figA}.  (We set $\hbar=1$ throughout).

The degeneracies of $H(0)$ are of course due to the enhanced rotational symmetry: without the last term in \HapperH, $H(b)$ commutes also with $J_x$ and $J_y$, so the eigenvectors organize into multiplets of the total $SU(2)$ algebra generated by $\vec{J}$, with $H(b)$ acting as a multiple of the identity within each multiplet.  Non-abelian symmetry is indeed the usual way of explaining degeneracies in simple quantum mechanical systems.  Despite some effort \cite{HapperUnpublished,BGX,YHA,BradleyJP}, simple expressions for the generators of a non-abelian symmetry group commuting with $H(k+1/2)$ have not come to light.

In \cite{HapperEtAl}, the Hamiltonian \HapperH, augmented by a spin-axis interaction, was used to describe the dynamics of the total nuclear spin $\vec{K}$ and the total electronic spin $\vec{S}$ of ${\rm Rb}_2$ molecules in an electronic spin triplet state, subjected to a magnetic field.  Resonances were observed in the spin relaxation rate, as a function of magnetic field, near the special values of the magnetic field corresponding to $q=k+1/2$.  The exact degeneracy for the Hamiltonian \HapperH\ was then used to argue that the added spin-axis interactions are a necessary and important part of spin relaxation.

The degeneracy of $H(k+1/2)$ can be demonstrated through direct calculation.  Observe first that because $[J_z,H(b)]=0$, $H(b)$ is block diagonal in a basis where $J_z$ is diagonal; and because $s=1$, the blocks are at most $3 \times 3$.  For $|m|<k$, the $J_z = m$ block of $H(k+1/2)$ is
 \begin{widetext}
 \eqn{mBlock}{
  H_m = \begin{pmatrix} k+m & \sqrt{(k+m)(k-m+1)/2} & 0  \cr
           \sqrt{(k+m)(k-m+1)/2} & 1/2 & \sqrt{(k-m)(k+m+1)/2}  \cr
           0 & \sqrt{(k-m)(k+m+1)/2} & -k-m-1 \end{pmatrix} \,,
 }
 \end{widetext}
in a basis where $S_z = 1$ in the first row, $0$ in the second, and $-1$ in the third.  The determinant of each of these blocks vanishes, and also for the smaller blocks for $|m| \geq k$, except when $J_z = k+1$ or $J_z = -k$.  Further calculation shows that each block with zero determinant has exactly one eigenvector with zero eigenvalue.  This demonstration is due to Happer \cite{HapperUnpublished}, who went on to show, based on an analysis of the secular equation for $H_m$, that
 \eqn{PH}{
  P_H = 1 - {H(k+1/2) [2 H(k+1/2) + 1] \over (2k+1) (2J_z + 2k + 1)}
 }
is a projection operator whose image is the kernel of $H(k+1/2)$.  The image of a linear operator $O$ is the set of all vectors $|u\rangle$ which can be written in the form $|u\rangle = O|v\rangle$, while the kernel of $O$ is the set of all vectors $|v\rangle$ such that $O|v\rangle = 0$.

In \cite{YHA} it was observed that $P_H O P_H$ commutes with $H(k+1/2)$ for any operator $O$ (this fact depends only on having the image of $P_H$ fall in a single eigenspace of $H(k+1/2)$), and based on this observation symmetry operators were constructed which generate an $SU(2)$ algebra for which the $(2k+1)$-dimensional kernel of $H(k+1/2)$ furnishes the spin $k$ representation.  The final expression for the generators is somewhat complicated, and its derivation depends crucially on the detailed knowledge of the spectrum that one can extract from \mBlock.

We therefore seek alternative approaches to understanding the mysterious degeneracy of $H(k+1/2)$.  What special feature of $H(k+1/2)$ makes a third of its eigenvalues vanish?  Why is the degeneracy only approximate for larger integer values of $s$?  What related Hamiltonians exhibit similar degeneracies, and why?  In section~\ref{Squares}, we will give partial answers to the first two of these questions based on crisp answers to the third.  In section~\ref{Higher}, we will give some preliminary indications that some of the special properties for $s=1$ extend in an approximate way to integer $s>1$.

\section{Squares of the basic Hamiltonian}
\label{Squares}

Consider the operators
 \eqn{SquareOps}{
  Z(b) &= H(b) H(-b) = X(b) + i b Y  \cr
  X(b) &= {1 \over 2} \left\{ H(b), H(-b) \right\}
   = (\vec{K} \cdot \vec{S} + 1/2)^2 - b^2 S_z^2  \cr
  Y &= {1 \over 2ib} [ H(b),H(-q) ] 
   = {1 \over i} [S_z, \vec{K} \cdot \vec{S} ]  \cr
   &= K_x S_y - K_y S_x \,.
 }
The relation between $X(b)$ and $H(b)$ is reminiscent of the relation between the energy and the supercharge in a supersymmetric theory: $2E = \{ Q,Q^\dagger \}$.  One may show that the quantum states annihilated by $E$ are precisely those annihilated by $Q$ and $Q^\dagger$: for any state $|v\rangle$,
 \eqn{SUSYarg}{
  \langle v | 2E | v \rangle = 
   \langle v | Q Q^\dagger | v \rangle + 
   \langle v | Q^\dagger Q | v \rangle \,,
 }
and upon observing that the terms on the right hand side are, respectively, the squares of the norms of $Q^\dagger|v\rangle$ and $Q|v\rangle$, the desired conclusion follows immediately.  But $H(-q)$ is not the adjoint of $H(b)$, so an analogous argument cannot be made in the present case \footnote{We could invent a new anti-linear involution on operators that carries $H(b)$ to $H(-b)$: for example, $O \to e^{i\pi J_y} O^\dagger e^{-i\pi J_y}$ has this property.  The associated inner product isn't positive definite, so the standard argument still fails.}.  Nevertheless, as we will argue below, the kernel of $X(k+1/2)$ is closely related to the kernel of $H(k+1/2)$ when $s=1$.

The properties of $X(b)$ are easy to understand because $(\vec{K} \cdot \vec{S} + 1/2)^2$ and $S_z^2$ each has only two distinct eigenvalues.  More particularly, there are $4k+2$ linearly independent vectors $|u_n\rangle$ with $(\vec{K} \cdot \vec{S} + 1/2)^2 |u_n\rangle = (k+1/2)^2 |u_n\rangle$, and likewise there are $4k+2$ linearly independent vectors $u_m$ with $S_z^2 |v_m\rangle = |v_m\rangle$.  The total Hilbert space is only $(6k+3)$-dimensional, so the $|u_n\rangle$ and the $|v_m\rangle$ can't all be linearly independent: at least $2k+1$ of the $|u_n\rangle$ and $|v_m\rangle$ can be chosen to coincide.  Another way to say this is that the subspaces $U$ and $V$ spanned by $\{ |u_n\rangle \}$ and $\{ |v_m\rangle \}$ must intersect over a subspace of dimension at least $2k+1$.  If we took $k=0$, this would amount roughly to the familiar statement that two planes in three dimensions intersect at least over a line.  Any vector $|w\rangle$ in the intersection $U \cap V$ is clearly an eigenvector of $X(b)$ with eigenvalue $(k+1/2)^2-b^2$.

A concise description of $U$ is that its orthogonal complement is the space where the total angular momentum $j$ is equal to $k$.  And the orthogonal complement of $V$ is the space where $S_z = 0$.

We conclude that $X(b)$ has a large degenerate eigenspace (at least $(2k+1)$-dimensional) for {\it any} value of $q$.  That eigenspace, constructed as $U \cap V$ in the previous paragraph, doesn't depend on $q$, but its $X(b)$ eigenvalue does; and that eigenvalue is $0$ when $b=k+1/2$.  In fact, the dimension of $U \cap V$ is exactly $2k+1$.  We would regard it as coincidence if it were bigger, just as it would be a coincidence if two planes chosen at random in three dimensions happened to be parallel.  By the same arguments, a Hamiltonian constructed as a sum of two terms which fail to commute and have eigenspaces with dimensions adding up to $r$ plus the dimension of the total Hilbert space must itself have an eigenspace with degeneracy of at least $r$---and the generic situation is for the degeneracy to be precisely $r$.  Generalizing to more than two terms is straightforward, but then the degeneracies for the individual terms must be quite large for the total Hamiltonian to have a degeneracy.

What does this have to do with the original problem of degeneracies of $H(k+1/2)$?  Suppose $|w\rangle$ is in the kernel of $X(k+1/2)$---that is, the space $U \cap V$ described above.  Then
 \eqn{HHonW}{
  H(k+1/2) H(-k-1/2) |w\rangle &= Z(k+1/2) |w\rangle  \cr
   &= i (k+1/2) Y |w\rangle \,,
 }
so if it happened that $Y|w\rangle=0$, we would conclude that the vector $H(-k-1/2) |w\rangle$ is one of the degenerate eigenvectors of $H(k+1/2)$.  Amazingly, $Y|w\rangle=0$ precisely for $|w\rangle \in U \cap V$.  If we just assume this special property of $Y$, the original problem is almost solved: acting with $H(-k-1/2)$ on each of the $2k+1$ vectors in $U \cap V$ produces $2k+1$ vectors annihilated by $H(k+1/2)$.

The only way this line of reasoning could fail is if some $|w\rangle \in U \cap V$ is annihilated by $H(-k-1/2)$.  Suppose this happens for $p$ vectors: more precisely, suppose in an orthonormal basis for $U \cap V$, $p$ vectors are annihilated by $H(-k-1/2)$ and the others are carried by it to linearly independent images.  Then we would have demonstrated that $H(-k-1/2)$ annihilates at least $p$ vectors while $H(k+1/2)$ annihilates at least $2k+1-p$.  But $H(b)$ and $H(-b)$ obviously have the same spectrum, so we have shown that the number of independent vectors annihilated by $H(k+1/2)$ is at least the larger of $p$ and $2k+1-p$---thus at least $k+1/2$ vectors.

It would be generic to have $p=0$, in the sense that if we knew only that $H(-k-1/2)$ and $X(k+1/2)$ failed to commute, we would guess that their eigenspaces did not intersect except at the origin.  If this were so, then we would indeed have constructed the full $(2k+1)$-dimensional degenerate eigenspace of $H(k+1/2)$ by acting with $H(-k-1/2)$ on the kernel of $X(k+1/2)$.  The situation is not quite so simple: the highest weight eigenvector of $J_z$, whose eigenvalue is $k+1$, is annihilated by both $H(-k-1/2)$ and $X(k+1/2)$.  With no further constraints, genericity suggests that $p=1$ and that the construction explained after \HHonW\ gives us $2k$ of the $2k+1$ degenerate eigenvectors of $H(k+1/2)$.  This is right: the only zero eigenstate of $H(k+1/2)$ that we miss is the one with $J_z = -k+1$.

We still have to show that $Y|w\rangle=0$ for $|w\rangle \in U \cap V$.  In investigating the properties of $Y$ we encounter a pleasant surprise: for any $s \leq k$, $Y$ annihilates $2k+1$ linearly independent vectors if $s$ is an integer, and $2s+1$ vectors if $s$ is a half-integer.  To see this, consider a basis in which $J_z$, $K_z$, and $S_z$ are all diagonal, arranged so that $Y$ is block diagonal in blocks with definite $J_z$ eigenvalues $m$.  Recall from the theory of addition of angular momenta that the $|m|=k+s$ blocks are one-dimensional, the $|m|=k+s-1$ blocks are two-dimensional, and so forth until we reach the blocks with $|m| \leq k-s$, which are all $(2s+1)$-dimensional.  Within each block, let us arrange the basis so that the $S_z$ eigenvalue decreases as one descends through the rows.  Noting that
 \eqn{Yform}{
  Y = {i \over 2} (K_+ S_- - K_- S_+) \,,
 }
where $K_\pm = K_x \pm i K_y$ and $S_\pm = S_x \pm i S_y$, it is apparent that within each block, the only non-zero entries are one step off the diagonal.  The determinant of each block must therefore vanish when the dimension of the block is odd: this owes to the fact that there is no permutation of an odd number of objects that replaces each object by one of its nearest neighbors.  Counting up the number of odd-dimensional blocks leads us immediately to the conclusion that the dimension of the kernel of $Y$ is at least $2k+1$ if $2s+1$ is odd, and at least $2s+1$ if $2s+1$ is even.

In the same basis, with phases chosen so that the matrix elements of $K_+$ and $S_+$ are all positive, it is straightforward to show that $Y$ annihilates the vector in the odd-dimensional block with $J_z=m$ whose first component is $1$ and whose subsequent components are alternately $0$ and
 \begin{widetext}
 \eqn{OddComponents}{
  \prod_{m_s < \ell \leq m_s^{\rm max} \atop \ell-s\ \rm even} 
     \sqrt{(k-m+\ell)(k+m-\ell+1)(s+\ell)(s-\ell+1) \over
           (k-m+\ell-1)(k+m-\ell+2)(s+\ell-1)(s-\ell+2)} \,,
 }
 \end{widetext}
where $m_s$ is the $S_z$ eigenvalue of the row in question, and $m_s^{\rm max}$ is the maximum eigenvalue of $S_z$ within the block (which is $s$ except when $m<-k+s$).  To demonstrate \OddComponents, one needs relations like $K_+ |m_k\rangle = \sqrt{(k-m_k)(k+m_k+1)} |m_k+1\rangle$ for unit eigenvectors of $K_z$.

Clearly, the vectors described around \OddComponents\ are orthogonal to all $S_z=m_s$ eigenspaces with $s-m_s$ odd.  In particular, for $s=1$, they are orthogonal to the $S_z=0$ eigenspace.  They also are orthogonal to the $j=k$ eigenspace.  This fact may seem intuitive, since in the $j=k$ eigenspace the spin $\vec{S}$ is in some rough sense orthogonal to the spin $\vec{K}$ (so it would be surprising for the cross-product $\vec{K} \times \vec{S}$ to vanish); but the only proof we can offer is based on straightforwardly showing that the $j=k$ eigenvectors in the three-dimensional blocks with $J_z=m$ are
 \eqn{gkEvs}{
  \begin{pmatrix}\sqrt{(k+m)(k-m+1)} \cr
                 -\sqrt{2} m \cr 
                 -\sqrt{(k-m)(k+m+1)}\end{pmatrix} \,,
 }
which is easily seen to be orthogonal to the vector described around \OddComponents.  Thus indeed $Y|w\rangle=0$ for $|w\rangle \in U \cap V$.

\section{Higher values of $s$}
\label{Higher}

For integer $s>1$, and for $k>s$, the spectrum of $H(b)$ exhibits an approximate $(2k+1-s)$-fold degeneracy for a value $b_*$ of $b$ slightly smaller than $k+1/2$ \footnote{For half-integer $s$, no such degeneracy appears: in fact, eigenvalues of $H(b)$ never cluster close to zero, heuristically because half-integer $s$ means that $\vec{S}$ cannot be nearly orthogonal to any given vector---in particular, to $\vec{K} - b \hat{z}$.}.   This is clearly a similar phenomenon to the exact degeneracy for $s=1$.  But $H(b)$ seems altogether more complicated for $s>1$: it has many two-fold degeneracies, seldom if ever at the same value of $b$; whereas for $s=1$, the only degeneracies are the large ones at $b=0$ and $|b|=k+1/2$.  This suggests that some notion of integrability applies to the $s=1$ case but not the $s>1$ cases.

There are nevertheless some hints that our methods have some applicability to integer $s>1$.  Define 
 \eqn{HXDefs}{
  H_s(b) &= \vec{K} \cdot \vec{S} + b S_z + r_s  \cr
  X_s(b) &= {1 \over 2} \{ H_s(b), H_s(-b) \} = 
   (\vec{K} \cdot \vec{S} + r_s)^2 - b^2 S_z^2  \cr
  Y &= {1 \over 2ib} [ H_s(b), H_s(-b) ] = K_x S_y - K_y S_x \,,
 }
with $r_s$ chosen to put the approximate degeneracy near zero eigenvalue: $r_s = s(s+1)/4$ will do at least through $s=5$.  Now $X_s(b_*)$ also has $2k+1-s$ eigenvectors with eigenvalue close to zero.  For $s=2$, in all but one of the $2k-4$ eigenspaces of $J_z$ where $Y$, $X_2(b_*)$, and $H_2(b_*)$ all have nearly a zero eigenvector, these three eigenvectors are almost linearly dependent---to within a few percent for $k=6$ and $7$.  And for $s=3$, acting with $H(-b_*)$ upon the $2k+1$ eigenvectors of $X_3(b_*)$ with the smallest eigenvalues leads almost to $2k-3$ of the nearly degenerate eigenvectors of $H_3(b_*)$---to an accuracy such that the dot product between unit vectors, one chosen arbitrarily from the second, smaller set and the other chosen carefully from the first, is always between $0.9$ and $0.95$, for $k=7$ and $8$.  This last observation is particularly striking: the construction we gave following \HHonW\ almost works, even though the zero eigenvectors of $Y$ are far from the nearly degenerate eigenvectors of $X_3(b_*)$.

Of course, the ideal would be to find some perturbation of $H_s(b)$ that has an exact degeneracy.  Then the approximate degeneracy might be understood via first order perturbation theory.  An elementary example of how this can work is furnished by $H(q,p) = \vec{K} \cdot \vec{S} + q S_z + p K_z$, which for $q$ and $p$ small has $(2j+1)$-fold approximate degeneracies among eigenvectors which nearly have total angular momentum $j$, provided (in the terminology of the Wigner-Eckart theorem) the reduced matrix element of $q \vec{S} + p \vec{K}$ vanishes: that is, $q [ j(j+1) + s(s+1) - k(k+1) ] + p [ j(j+1) + k(k+1) - s(s+1) ] = 0$.  For $q$ and $p$ solving this last equation.

\section{Conclusions}
\label{Conclude}

We have advanced general arguments which account for the large degeneracy of eigenvalues of the Hamiltonian $H(k+1/2)$ for $s=1$.  The key idea, inspired by supersymmetry \footnote{Our construction should not be confused with supersymmetric quantum mechanics \cite{WittenSUSYQ}, in which a pair of Hamiltonians is expressed in terms of a certain anti-commutator.}, was to relate the kernel of the anti-commutator of two operators to the kernels of the individual operators.  Our construction actually guarantees only a $(k+1/2)$-fold degeneracy (see the discussion below \HHonW).  However, the observed $(2k+1)$-fold degeneracy has a satisfying explanation within our construction, which we can restate as follows: we start with operators $H(k+1/2)$ and $H(-k-1/2)$ whose commutator and anti-commutator obviously annihilate $2k+1$ independent quantum states.  If two such operators had no other special properties, then the argument following \HHonW\ would tell us that $H(k+1/2)$ itself has $2k+1$ zero-energy eigenstates---the desired conclusion.  But $H(k+1/2)$ and $H(-k-1/2)$ {\it do} have an additional special property, having to do with a common eigenstate for $H(-k-1/2)$ and $\{H(k+1/2,H(-k-1/2)\}$.  This just means that of the $2k+1$ zero-energy eigenstates, all but one is constructed as indicated in the discussion following \HHonW, and the last one is special.

To recap: the basic question we set out to answer was, what is special about the original Hamiltonians $H(b)$ and $H(-b)$ that makes a third of their eigenvalues vanish when the magnetic field parameter $b$ takes on a special value?  The answer, modulo the one exceptional eigenstate, is that the commutator and anti-commutator of these Hamiltonians have obvious large degeneracies that go to zero energy at the special value of $b$.

In the process of formulating these general arguments, we have noted a large class of Hamiltonians with large degeneracies but no obvious symmetries, namely ones which can be written as a sum of terms, each with a large degeneracy, which don't commute with one another.  We have also noted a special operator, $Y = K_x S_y - K_y S_x$, which has a large kernel for any $k$ and $s$.

The observations that our construction of the zero eigenvalues of $H(k+1/2)$ extends to an approximate treatment for $s=3$, and that for $s=2$ there is a wholly different but non-trivial relationship among the eigenspaces of $H_2(b_*)$ and the ancillary commutator and anti-commutator operators, suggest that much more remains to be learned about this class of Hamiltonians.

\section*{Acknowledgements}

We thank W.~Happer for useful discussions.  The work of S.S.G.~was supported in part by the Department of Energy under Grant No.\ DE-FG02-91ER40671, and by the Sloan Foundation.

\bibliography{kernel}

\end{document}